\documentclass[12pt]{article}

\usepackage{subcaption}
\usepackage{graphicx}
\usepackage{times}
\usepackage{mathtools,amssymb}
\usepackage{lineno}
\newenvironment{sciabstract}{%
\begin{quote} \bf}
{\end{quote}}

\title{Stable radiation field positron acceleration in a micro-tube}
\author
{Meiyu Si,$^{1,2,3}$ Yongsheng Huang,$^{2\ast}$ Manqi Ruan,$^{1}$ Baifei Shen,$^{4}$ Zhangli Xu,$^{4}$ \\
Tongpu Yu,$^{5}$ Xiongfei Wang,$^{6,7}$ Yuan Chen,$^{8}$\\
\\
\normalsize{$^{1}$Institute of High Energy Physics, Chinese Academy of Sciences, Beijing 100049, China}\\
\normalsize{$^{2}$School of Science, Shenzhen Campus of Sun Yat-sen University, Shenzhen 518107, China}\\
\normalsize{$^{3}$University of Chinese Academy of Sciences, Beijing 100049, China}\\
\normalsize{$^{4}$Department of Physics, Shanghai Normal University, Shanghai 200234, China}\\
\normalsize{$^{5}$Department of Physics, National University of Defense Technology, Changsha 410073, China}\\
\normalsize{$^{6}$School of Physical Science and Technology, Lanzhou University, Lanzhou 730000, China}\\
\normalsize{$^{7}$Lanzhou Center for Theoretical Physics, Laboratory of Theoretical Physics of Gansu Province}\\
\normalsize{and Frontiers Science Center for Rare Isotopes, Lanzhou University, Lanzhou 730000, China}\\
\normalsize{$^{8}$The Institute for Advanced Studies of Wuhan University, 299, Bayi Road, Wuhan 430072, China}\\
\\
\normalsize{$^\ast$ E-mail:  huangysh59@mail.sysu.edu.cn.}
}

\date{}

\begin{document} 


\baselineskip24pt


\maketitle 


\begin{sciabstract}
Nowadays, there is a desperate need for an ultra-acceleration-gradient method for antimatter particles, which holds great significance in exploring the origin of matter, CP violation, astrophysics, and medical physics. Compared to traditional accelerators with low gradients and a  limited acceleration region for positrons in laser-driven charge separation fields, we propose an innovative high-gradient  positron acceleration mechanism with implementation advantages. Injecting a relativistic electron beam into a dense plasma micro-tube generates a stable and periodic high-intensity mid-infrared radiation (mid-IR) field, reaching tens of $\mathrm{GV/m}$. This field, propagating synchronously with the electron beam, achieves a 1 $\mathrm{GeV}$ energy gain for the positron bunch within 140 picoseconds with a minimal energy spread—approximately 1.56$\%$ during a stable phase. By utilizing continuous mid-IR, the efficiency of energy transfer from the electron beam to either a single positron bunch or three positron bunches simultaneously could reach up to 20$\%$ and 40$\%$, respectively. This acceleration scheme can achieve cascaded acceleration for a single positron bunch and series acceleration for multiple positron bunches in a continuous, stable, and efficient manner.
\end{sciabstract}

\section{Introduction}
The high-efficiency particle acceleration technology holds unique importance in astroparticle physics, medical physics, and material sciences\cite{L1979,G2014,H2006}. Traditional accelerators are limited by the ionization threshold of materials\cite{Gschwendtner,Jiao}, typically reaching gradients of around hundreds of $\mathrm{MV/m}$. Therefore, to obtain high-energy particle beams, larger accelerator facilities are required, leading to increased costs. Laser and particle beam-driven plasma charge separation wakefields can generate higher acceleration gradients, typically ranging from several tens to several hundreds of $\mathrm{GV/m}$, while allowing control over the energy spread and beam emittance of the accelerated particles\cite{Tajima,Katsouleas,Faure,Mangles,Geddes,WangJ,DaiYN}. Significant progress has been made in electron beam acceleration in this regard.  Simultaneously, various positron acceleration schemes are continuously being proposed. The fundamental plasma bubble acceleration, driven by a laser or an electron beam\cite{2016,Zhu,L2007,D2000,H2003,W2008,V2014}, results in positrons undergoing a shorter acceleration phase in comparison to electron plasma acceleration, which can lead to instability. Concurrently, strong transverse forces can also diminish the quality of the positron beam. 
Therefore, the concept of electron beam driving within a hollow plasma channel is proposed, aiming to achieve longitudinally stable acceleration for positrons\cite{Schroeder1999,Schroeder2013,Gessner2016,Luwei}. For proton beam driven acceleration, there is an increase in particle divergence due to the strong Coulomb force, with additional four-stage iron focusing guidance applied to the hollow channel\cite{Y2014}. When considering the driving of electron beams, the density of the hollow plasma channel has a different effect on positron acceleration. Positron acceleration in low-density channels occurs during the phase formed between the first and second bubbles. Addressing the stability of the acceleration phase and the preparation of the given-density plasma channel is challenged.

\noindent Ultra-short electron beams passing through the copper film target interface generate strong coherent transition radiation (CTR), which can be used to capture and accelerate positron beams produced through the Bethe-Heitler (BH) process in the target\cite{Xu2020}. Additionally, the issues of radiation and electron beam divergence are suggested to be solved by strengthening the magnetic field and using a proper waveguide\cite{Xu2023}. Simultaneously, single-cycle tunable infrared pulses generated in density-tailored plasma can be used to drive the acceleration of positrons\cite{ZhuXL,Si}. In this paper, we propose a novel mechanism based on the continuous emission of high-intensity mid-infrared radiation (mid-IR). This radiation is generated by the oscillations of surface-nanometer electron films driven by a relativistic electron beam in a plasma and/or metal micro-tube. This method allows for the construction of a highly efficient and stable uniform field structure. The stable wavelength of the periodic acceleration field is approximately 24 $\mathrm{\mu m}$ in the mid-IR range, and it can be tuned by adjusting the micro-tube radius, for example, into the terahertz range.  The acceleration field can reach several tens of $\mathrm{GV/m}$ and depends on the charge of the driving electron beam. For an electron beam with a charge of several tens of $\mathrm{nC}$, the acceleration field can even reach several hundreds of $\mathrm{GV/m}$, resulting in positron acceleration energy gains ranging from tens to hundreds of $\mathrm{GeV}$.

\noindent The self-consistent fields of a relativistic electron beam in a plasma micro-tube and the high-intensity continuous mid-IR generated by surface-nanometer film-plasma oscillations were analyzed. Field validation was performed by injecting a positron witness bunch. The results indicate that positrons can achieve energy gains at the GeV level within 140 $\mathrm{ps}$, with an achievable energy spread of up to 1.56$\%$ and an efficiency of up to 20$\%$. The quality of the positron beam can be adjusted by modifying the driving parameters, allowing for positron energies ranging from $\mathrm{GeV}$ to even hundreds of $\mathrm{GeV}$. In the uniform mid-IR acceleration field, the transverse field ($E_y-cB_z$) focuses the positron bunch within the first 21 $\mathrm{ps}$, followed by weak defocusing until the transverse field reduces to zero due to energy loss and beam break of the electron beam. Through several 2-dimensional particle-in-cell (PIC) simulations, the transverse spatial misalignment of positron injection was allowed to be a few microns if a 10$\%$ relative loss in energy transfer efficiency is acceptable, where the accelerated energy and energy spread are hardly affected. The design scheme for cascaded acceleration of a single positron bunch and series acceleration of multiple positron bunches is presented in the 'Discussion' section.

\section{Analysis of $e^+$ acceleration process}
\subsection{Self-fields of relativistic $e^-$ beam in a plasma micro-tube}
For a relativistic electron beam propagating in the $x$ direction in free space with a normalized velocity $\beta_0 = v_x/c$ and a density distribution $\rho_b$, the initial current is given as $J(x, t=0)=\beta_0 \rho_b$. The relativistic Poisson's equation in the laboratory frame can be written as \cite{Londrillo,Massimo,Marocchino}
\begin{equation}
\bigg[\frac{1}{\gamma^{2}_{0}}\frac{\partial^2}{\partial^2_{x^{2}}}+\frac{\partial^2}{\partial^2_{y^{2}}}+\frac{\partial^2}{\partial^2_{z^{2}}}\bigg]\Phi=-\rho_b,
\end{equation}
where $\gamma_0=1/\sqrt{1-\beta^{2}_{0}}$ is the Lorentz factor of the relativistic electron beam, and $\Phi$ is the electrostatic potential. To solve the initial potential $\Phi$ in free space, the standard fast Fourier transformation (FFT) algorithm is used to perform the cosine transform. Therefore, the electric fields of the initial electron beam can be computed as $E_x=-\frac{1}{\gamma_0^{2}}\partial_x\Phi, E_y=-\partial_y\Phi$. Due to the $\frac{1}{\gamma_0^2}$ factor, the longitudinal electric field $E_x$ has a much smaller magnitude than the transverse electric field $E_y$.

\noindent In this study, the initial electric field distribution was obtained through a PIC simulation for a Gaussian electron bunch with a charge of 3.402 $\mathrm{nC}$ and a Lorentz factor of $\gamma_0\approx1957$ corresponding to an electron beam energy of 1 $\mathrm{GeV}$. Detailed parameters can be found in the Methods section. The initial $E_x$ and $E_y$ of the relativistic electron beam are shown in Figs. 1(A) and (B), respectively, with their subplots displaying the generated electrostatic fields. In the case of a plasma micro-tube with a radius of 50 $\mathrm{\mu m}$, the $E_y$ of the relativistic electron beam was applied to the inner walls due to its larger value (approximately 200 $\mathrm{GV/m}$) compared to that of a static electron bunch. The field decays exponentially as it passes through the dense plasma film (now denoted as $E_{yp}$), and can be expressed as
\begin{equation}
\begin{split}
    E_{yp}=E_0e^{-\sqrt{k_x^2-1+\omega^2_{pe}}\delta y},\\
    \delta y=y'-r,
\end{split}
\end{equation}
\noindent where $r=\sqrt{y^2+z^2}$ is the radius of the plasma micro-tube, $y'$ is the penetration depth of the electric field in the plasma, and $\delta y$ is the charge separation distance. The plasma frequency was calculated as $\omega_{pe}=\sqrt{4\pi e^2 \rho/m}$, which is approximately $5.6\times10^{15} \mathrm{Hz}$, where $\rho$ is the density of dense plasma micro-tube. The longitudinal length of the electron beam was fixed at $\lambda$ $\sim7$ $\mathrm{\mu m}$, and the normalization parameter was set as $L=c/\omega_L\times10^6$ $\mathrm{\mu m}$, where $\omega_L=2\pi c/\lambda$. A strong charge separation field $E_{cs}$ proportional to the separation distance $\delta y$, expressed as
\begin{equation}
 E_{cs}=\frac{e}{\varepsilon_0}\rho\delta y.
\end{equation}

\noindent Fig. 1(C) displays the distribution of $E_{yp}$ within a dense plasma micro-tube with a radius of $r=$ 50 $\mathrm{\mu m}$. The resultant force for this setup is given an $F = F_{yp} - F_{cs} = -e(E_{yp} - E_{cs})$. For $y>0$, the electron film was pushed back and forth between regions of $F>0$ and $F<0$ to realize the oscillation of the surface-nanometer film - plasma electrons. Fig. 1(D) displays the distribution of the electric field $E_{yp}$ and the charge separation field $E_{cs}$ within the dense plasma micro-tube. For the surface-nanometer electron film with $\delta_y=1.1$ $\mathrm{nm}$, the resultant force $F$ nullifies, and the plasma electrons gain the maximum momentum. 
\begin{figure}
	\centering
	\includegraphics[width=\columnwidth]{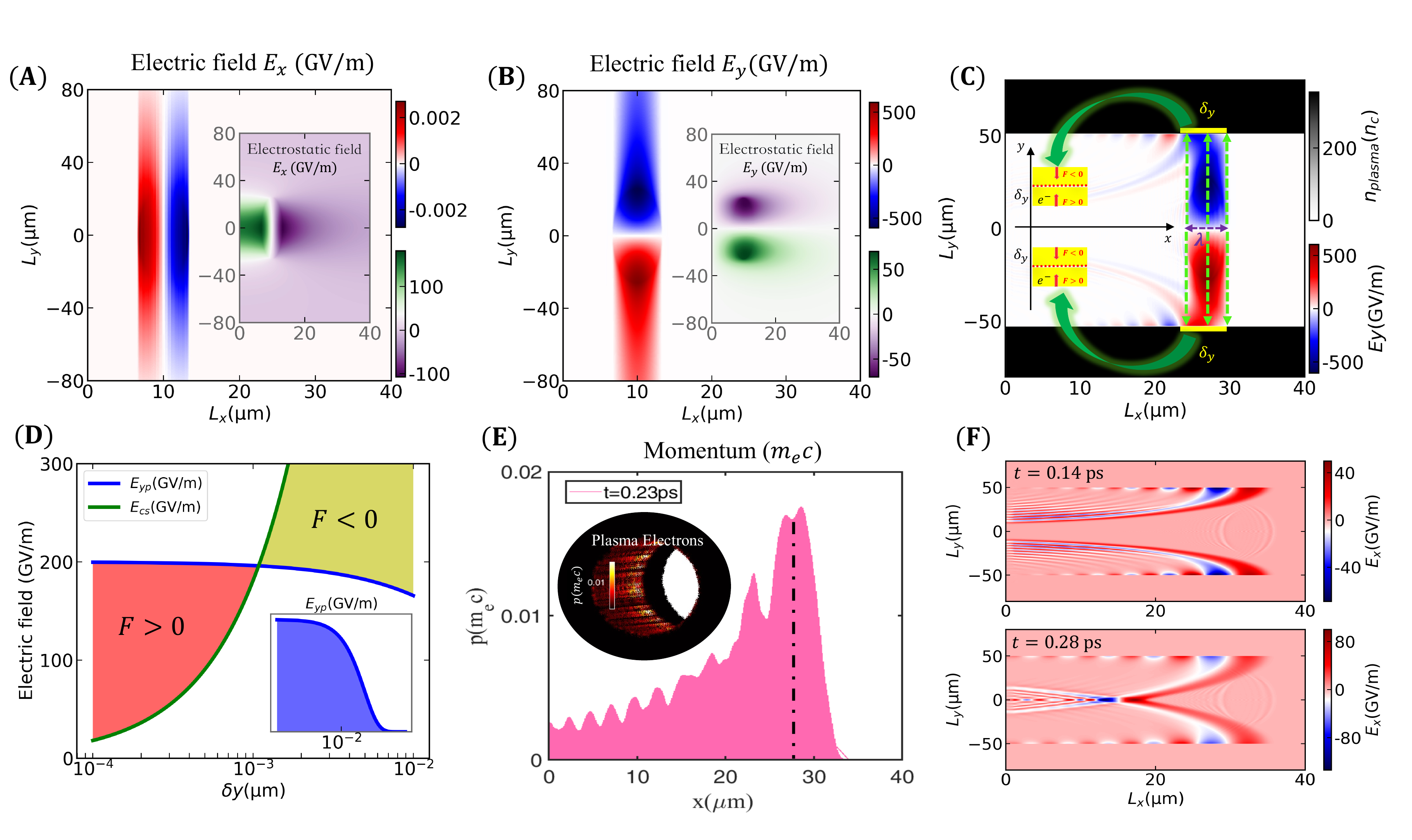}
	\caption{Self-consistent fields and the high-intensity continuous mid-IR acceleration field formed in the dense plasma micro-tube: ({\bf A}) The initial longitudinal electric field $E_x$, and ({\bf B}) the initial transverse electric field $E_y$ of the relativistic electron beam. The subplots display the electrostatic field of a static electron beam; ({\bf C}) The strong transverse field of a relativistic electron beam at the inner-surface of the plasma micro-tube; ({\bf D}) The trend of the resultant force $F$ of the electric field $E_{yp}$ in the dense plasma, and the charge separation field $E_{cs}$; ({\bf E}) The momentum distribution of the surface-nanometer plasma electron film. The maximum momentum of the plasma, obtained as $\sim$0.016 $\mathrm{m_ec}$, corresponds to the region where the driven beam is located; and ({\bf F}) The acceleration (radiation) field distribution in the plasma micro-tube with a radius of $50$ $\mathrm{\mu m}$ at $t=0.14$ $ps$ and $t=0.28$ $ps$.}
	\label{Figure2}
\end{figure}

\subsection{High-intensity mid-IR acceleration field}
The resultant force $F$ applied by the electric field $E_{yp}$ and the charge-separation field $E_{cs}$ within the dense plasma causes oscillations of the surface-nanometer electron film, moving in and out of the inner-surface. The energy of these electrons can be obtained as

\begin{equation}
\int_{0}^{\delta_y}{(eE_{yp}-eE_{cs}})dy=(\gamma_p -1)m_ec^2,
\end{equation}

\noindent where $\gamma_p$ is the Lorentz factor of the plasma electrons, and $m_e$ is electron mass. For $\delta y=1.1$ $\mathrm{nm}$, $\gamma_p$ is computed to be $\sim$1.0002, and the normalized electron velocity in the plasma, $\beta_p$, is computed to be 0.0206. Fig. 1(E) shows the momentum distribution of the oscillating electrons from the PIC simulation. Only the electrons on the surface-nanometer film gain kinetic energy. As the electron beam passes through, the oscillations weaken, and the velocity decreases. Initially, the surface-nanometer electron film starts to oscillate and emit continuous mid-IR radiation. The electric field radiated through the oscillations of an electron, $E_{rad}$, is then obtained as

\begin{equation}
E_{rad}=\frac{e}{4\pi \varepsilon_0}\bigg[\frac{\vec{n}\times(\vec{n}\times \dot{\vec{\beta_p}})}{Rc(1-\beta_p\cos\theta)^3}\bigg],
\end{equation}
\noindent where $\dot{\vec{\beta_p}}=e\vec E/(cm_e\gamma_p^3)$ and $\theta$ is the angle between the solid angle direction $\vec{n}$ and $\vec{\beta_p}$. For the interaction region, the total electric field $E_{rad}^{tot}$ is then calculated through the 3-dimensional integration with the micro-tube density as
\begin{equation}
E_{rad}^{tot}=\int E_{x,rad}\rho xd\varphi dxdy= \frac{e\dot{\beta_p}\rho\delta_y p}{2\varepsilon _0c} \int_{0}^{\theta}\frac{\sqrt{1-\cos^2\theta} \sin^2\theta}{(1-\beta_p\cos\theta)^3\cos^2\theta}d\theta,
\end{equation}
\noindent where $E_{x,rad}=E_{rad}\sin\theta$ is the longitudinal radiation field, $p=R\cos\theta$ is the distance from the reference point to the micro-tube wall, and $L_y=50$ $\mathrm{\mu m}$. 

\noindent Using Eq. (6), the total radiation field was obtained for different parameter ranges. For example, with $p=1$ $\mathrm{\mu m}$, we obtained $E_{rad}^{tot}\approx38$ $\mathrm{GV/m}$. Fig. 1(F) shows the radiation field distribution in the plasma micro-tube obtained from the PIC simulation at $t=0.14$ $ps$ and $t=0.28$ $ps$, respectively. The radiation field produced by the oscillations of the electron film propagates from the inner wall to the center to catch up with the driving electron beam. Therefore, in Fig. 2(A), a high intensity acceleration field is formed in a stable TM mode with a mid-IR wavelength of $\sim$24 $\mathrm{\mu m}$. The relevant characteristics of the mid-infrared field inside the micro-tube have been discussed in reference \cite{Si}, including radiation intensity, wavelength, micro-tube radius, and so on.

\subsection{Results of $e^+$ beam acceleration}
As the relativistic electron beam propagates in the micro-tube, the continuous mid-IR evolves into a high-intensity stable period structure of the longitudinal acceleration field in Fig. 2(A). Figs. 2(A1) and (A2) provide a magnified view of the first period region and the phase of the electrons and positrons. The acceleration of the positron bunch in the dense plasma micro-tube was observed and analyzed through PIC simulations.

We analyzed the relationship between positron acceleration time, positron energy, and relative energy spread in Fig. 2(B). The positron bunch gains an energy increase of 1 $\mathrm{GeV}$ within 140 picoseconds, with a relative energy spread of $1.56\%$. Fig. 2(C) shows the decrease in the total charge of the electron beam and two cases of positron bunches (with initial energies of $1$ $\mathrm{GeV}$ and $5$ $\mathrm{GeV}$, both having a beam charge of $0.111$ ${\mathrm{nC}}$) during the acceleration process. In Fig. 2(D), the final energy spectrum of positron bunches with different initial energies and the electron beam at $t=140$ $\mathrm{ps}$ is displayed.
  
The energy transfer efficiency from the driving electron beam to the positron bunch is defined as $\eta = (\delta E_p \cdot Q_p) / (\delta E_e \cdot Q_e)$, where $\delta E_{e/p}$ represents the change in the energy of the electron or positron bunch, and $Q_{e/p}$ is the charge of the electron beam or positron bunch. The energy transfer efficiency was approximately $4\%$ and $5\%$ for positron bunches with initial energies of $1$ GeV and $5$ GeV, respectively. The energy transfer efficiency, $\eta$, increased with the initial charge of the positron bunch, $n_p$, and even reached up to $50\%$ for $n_p=1.85$ nC, as shown in Fig. 2(E). 

The acceleration field exhibits a stable period distribution, with the field intensity and stabilization time being correlated with the charge and energy of the driving electron beam. We can adjust the parameters of the injected electron and positron beams to meet various application requirements: (1) A positron beam with an initial energy of 50 $\mathrm{MeV}$ can be accelerated to several $\mathrm{GeV}$ within 140 $\mathrm{ps}$ using a driving electron beam with 1 $\mathrm{GeV}$ and 3 $\mathrm{nC}$. (2) The acceleration mechanism allows for cascaded acceleration, enabling the positron beam to achieve higher energy levels under less demanding initial conditions. (3) Positrons undergoing acceleration over a one-meter with a 45 $\mathrm{GeV}$ driving electron beam can gain an energy increase of 100 $\mathrm{GeV}$. 
\begin{figure}
	\centering
	\includegraphics[width=\columnwidth]{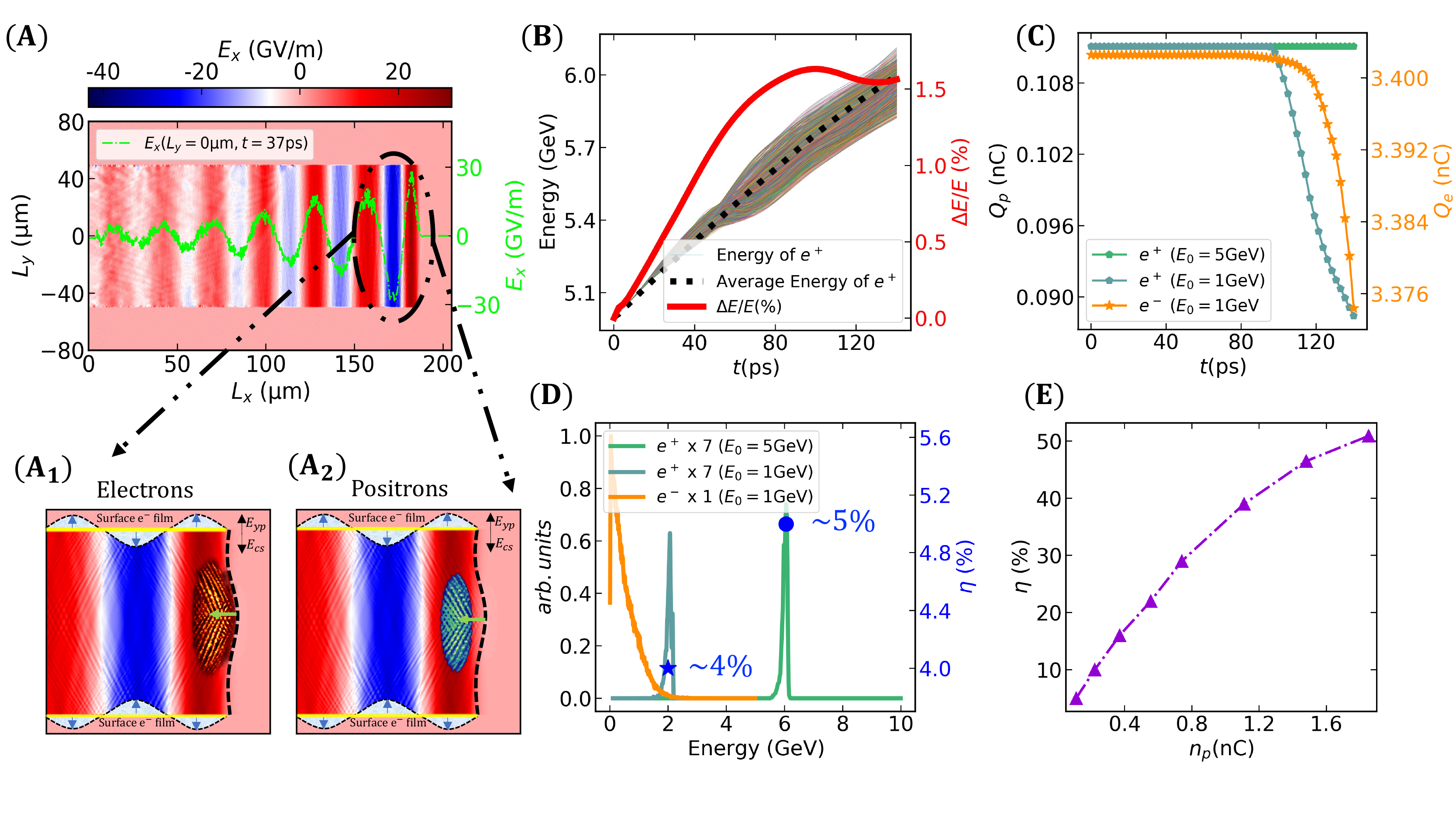}
	\caption{A novel scheme for positron bunch acceleration in a micro-tube: ({\bf A}) The periodically distributed mid-IR acceleration field is generated by the relativistic electron beam propagation within the dense plasma micro-tube. ({\bf B}) The relationship between the acceleration time and the positron energy, and the relative energy spread. At the acceleration distance of $4.2$ $\mathrm{cm}$, the positron bunch gains an average energy of $1$ $\mathrm{GeV}$. The relative energy spread of the positron bunch was to be $1.56\%$; ({\bf C}) The decrease in the total charge of the electron beam $Q_e$ and positron bunch $Q_p$ during acceleration process; ({\bf D}) The final energy spectrum of the positron bunches with different initial energies and the electron beam at $t=140$ $\mathrm{ps}$; and ({\bf E}) The energy transfer efficiency $\eta$ increases with the initial charge of positron bunch, $n_p$, and reaches up to 50$\%$ at $n_p=1.85\mathrm{nC}$. }
	\label{Figure1}
\end{figure}

\subsection{Evolution of transverse electromagnetic fields}
The stable acceleration of the positron bunch is achieved using the high-intensity longitudinal acceleration field generated by the oscillation of surface-nanometer electron film in the plasma micro-tube. Simultaneously, the transverse self-field generated by the relativistic electron beam also provides a transverse focusing force to the positron bunch. The transverse fields in the y- and z-directions, specifically, $E_y-cB_z$ and $E_z+cB_y$, were symmetrically configured in the PIC simulation.

\noindent Fig. 3(A) displays the transverse field distribution of the positron bunch at $0.1$ $\mathrm{ps}$, $1.9$ $\mathrm{ps}$, $21.0$ $\mathrm{ps}$, and $93.0$ $\mathrm{ps}$. In the early stage of the longitudinal acceleration ($0<t<21$ $\mathrm{ps}$), the transverse force remains focused. After $21$ $\mathrm{ps}$,  the transverse field becomes uniform at the edges, and the positron bunch remains stable in the transverse direction for several tens of picoseconds. Beyond 21 $\mathrm{ps}$, the field gradually weakens and becomes weakly defocused as the energy of the electron beam is consumed. Fig. 3(B) shows the corresponding phase-space distribution of the positron bunch, corresponding to the transverse field distribution in Fig. 3(A). Therefore, the analysis of the transverse field distribution and the transverse motion of the positron bunch in the y-direction revealed an initial weak focusing stage, followed by a weak defocusing stage, until the magnitudes of the transverse fields reduced to zero.
\begin{figure}
	\centering
	\includegraphics[width=\columnwidth]{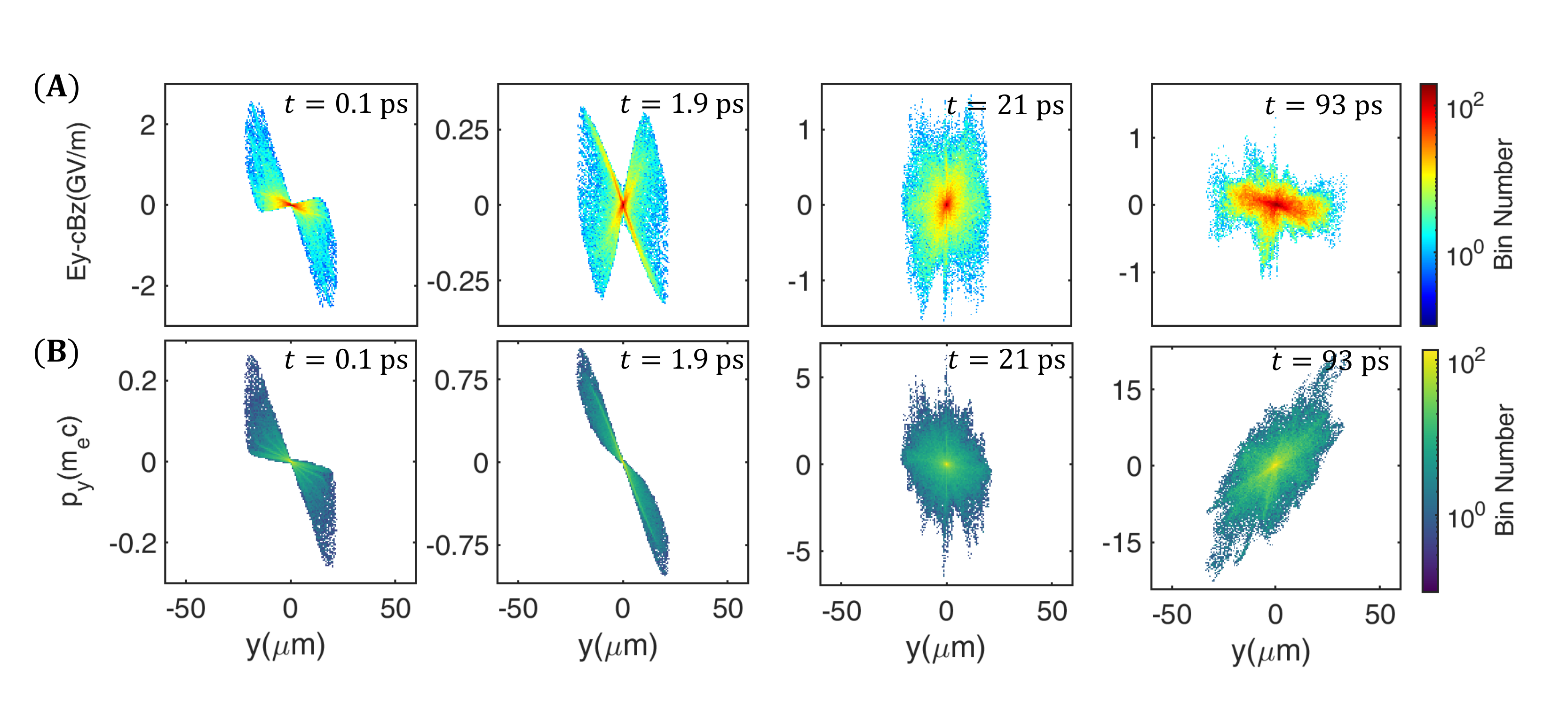}
	\caption{Distribution of the transverse field ($E_y-cB_z$) as experienced by the positron bunch while being accelerated in the plasma micro-tube: ({\bf A}) The transverse field distribution of positron bunch at $0.1$ $\mathrm{ps}$, $1.9$ $\mathrm{ps}$, $21.0$ $\mathrm{ps}$, and $93.0$ $\mathrm{ps}$, respectively; and ({\bf B}) The phase-space distribution of the positron bunch corresponding to the field distribution cases in ({\bf A}).}
	\label{Figure3}
\end{figure}

\subsection{Transverse off-axis $e^+$ bunch injection}
The proposed scheme enables stable and efficient acceleration of an externally injected positron bunch. Given the uniformity of the acceleration field, experimental misalignment of the injected positron bunch is inevitable but allowed within specific regions. To determine acceptable misalignment, several 2-dimensional PIC simulations were conducted with varying transverse injection deviations from the central axis. The simulations used the following key parameters: electron beam longitudinal and transverse lengths of $3.3$ $\mathrm{\mu m}$ and $26$ $\mathrm{\mu m}$, respectively, with a number density of $10^{24}$ $\mathrm{m^{-3}}$; positron bunch longitudinal and transverse lengths of $2$ $\mathrm{\mu m}$ and $22$ $\mathrm{\mu m}$, with a number density of 7$\times 10^{21}$ $\mathrm{m^{-3}}$; and initial injection positions of the positron bunch at $y_0=$ 0, 5$\mathrm{\mu m}$, and 10 $\mathrm{\mu m}$. Other simulation parameters were the same as those of the cylindrical coordinate simulations mentioned in the  Methods section. 
Fig. 4(A) shows the asymmetric phase-space distribution of a positron bunch with a $10$ $\mathrm{\mu m}$ positive y-axis deviation compared to the symmetric one. Fig. 4(B) shows the density distribution of the positron bunch in the $x-y$ plane after a long period of acceleration in both the asymmetric and symmetric cases. Due to the initial injection deviation in the y-direction, although the phase-space distribution of the positron bunch maintains this asymmetry during the acceleration process, the central energy and the energy spread of the accelerated positron bunch are hardly affected in the uniform longitudinal acceleration field. Fig. 4(C) displays the influence of the different transverse injection misalignments ($y_0=$ 0, 5$\mathrm{\mu m}$, and 10 $\mathrm{\mu m}$) on the accelerated energy spectrum of positron bunch. In each case, the positron bunch gained the same amount of energy at a given time $t$. However, the difference was the peak charge at the central energy which affects the energy transfer efficiency $\eta$ of the accelerated positron bunch; it decreases by $\sim10\%$ for the deviation of 10 $\mathrm{\mu m}$ relative to that without deviation. Therefore, the excellent tolerance for the injection deviation of the positron bunch greatly reduces the implementation difficulty of the experiment. 

\begin{figure}
	\centering
	\includegraphics[width=\columnwidth]{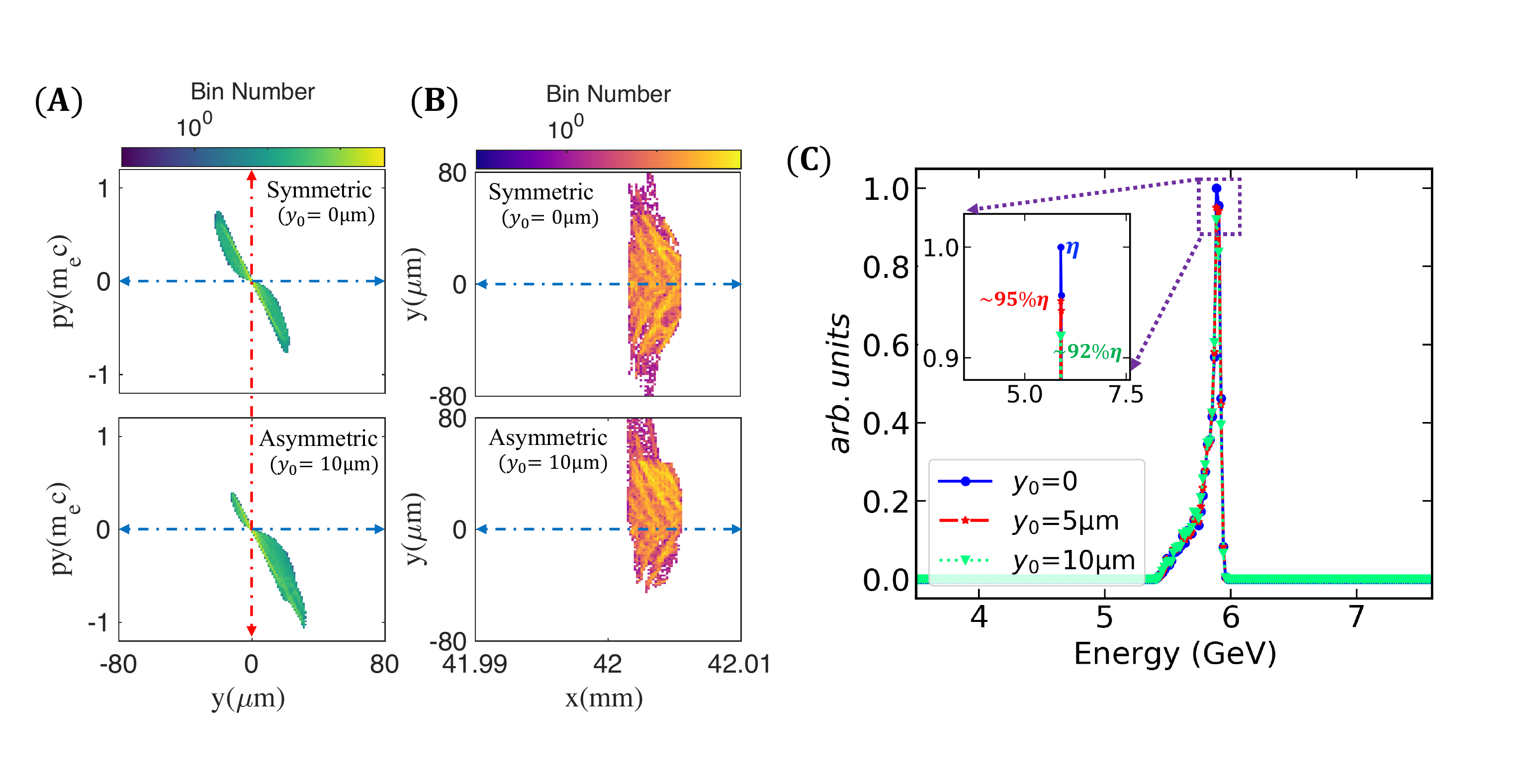}
	\caption{Impact of positron bunch injection deviation: ({\bf A}) The symmetric and asymmetric phase-space distribution of positron bunch; ({\bf B}) The symmetric and asymmetric density distribution of positron bunch in the $x-y$ plane after a long period of acceleration; and ({\bf C}) Effect of transverse injection misalignments at $y_0=$ 0, 5$\mathrm{\mu m}$, and 10 $\mathrm{\mu m}$, on the accelerated energy spectrum of positron bunch. $\eta$ is the energy transfer efficiency. }
	\label{Figure4}
\end{figure}

\section{Dicussions}
In our model, the density of the plasma micro-tube was fixed to be of the same order as the free-electron density in metals. Therefore, we propose a similar application of the method in metal micro-tubes. The transverse self-field $E_y$ of the relativistic electron beam (several hundreds of $\mathrm{GV/m}$) can directly ionize the surface of the metal micro-tube. It can also interact with the space-separation field to drive the oscillation of the ionized electron film, and the production of high-intensity, continuous mid-IR. This high-efficiency scheme has a high acceleration gradient and stable acceleration structure, and can also realize cascaded acceleration, allowing for a stable, continuous, and efficient positron acceleration. Fig. 5(A) displays the cascade design schematic diagram illustrating the acceleration of a positron bunch by electron beams, demonstrating a three-step acceleration process. The driving electron beam is extracted to the trash after completing a single acceleration. A cone-shaped coupling tube surrounded by a magnetic field. For a magnetic field of $B=1$ $\mathrm{T}$ and initial energy of 1 $\mathrm{GeV}$ of the electron or positron bunch, the deflection radius of particle bunch ($R=E/qvB$) was computed as $3.3$ $\mathrm{m}$. After setting the deflection angle to $\theta\approx1.8\times10^{-4}$ $\mathrm{rad}$ for $t=2$ $\mathrm{ps}$, the positron bunch was accelerated for $4.2$ $\mathrm{cm}$ in a mid-IR acceleration field through the micro-tube to obtain a $1$ $\mathrm{GeV}$ energy gain.
Due to the stable periodic structure of the mid-IR acceleration field in the micro-tube, the series acceleration can also be considered to obtain multiple high-energy positron bunches at the same time. This process is demonstrated by the 2-dimensional PIC simulation (Fig. 5(B)) of the acceleration field driven by a single electron beam in a micro-tube, accelerating three positron bunches, $x_{positrons}^{(1)}$, $x_{positrons}^{(2)}$, and $x_{positrons}^{(3)}$, simultaneously, starting from initial positions of $178$ $\mathrm{\mu m}$, $156$ $\mathrm{\mu m}$, and $126$ $\mathrm{\mu m}$, respectively. The final energy spectrum of the positron bunches is shown in Fig. 5(C) with the initial energy of $5$ $\mathrm{GeV}$. The energy transfer efficiency was found to be about 2-3 times higher than that of a single positron bunch acceleration. For an injected positron bunch with a long longitudinal dimension, such as 20 $\mathrm{\mu m}$, at a phase of $3\pi/2$, where the front and rear of the positron bunch are in an accelerating phase, and the center is in a decelerating phase. During long-distance propagation, the positrons in the central portion will be decelerated to the second acceleration phase. This allows the single bunch to split into two separate bunches for subsequent individual acceleration.

\begin{figure}
    \centering
    \includegraphics[width=\columnwidth]{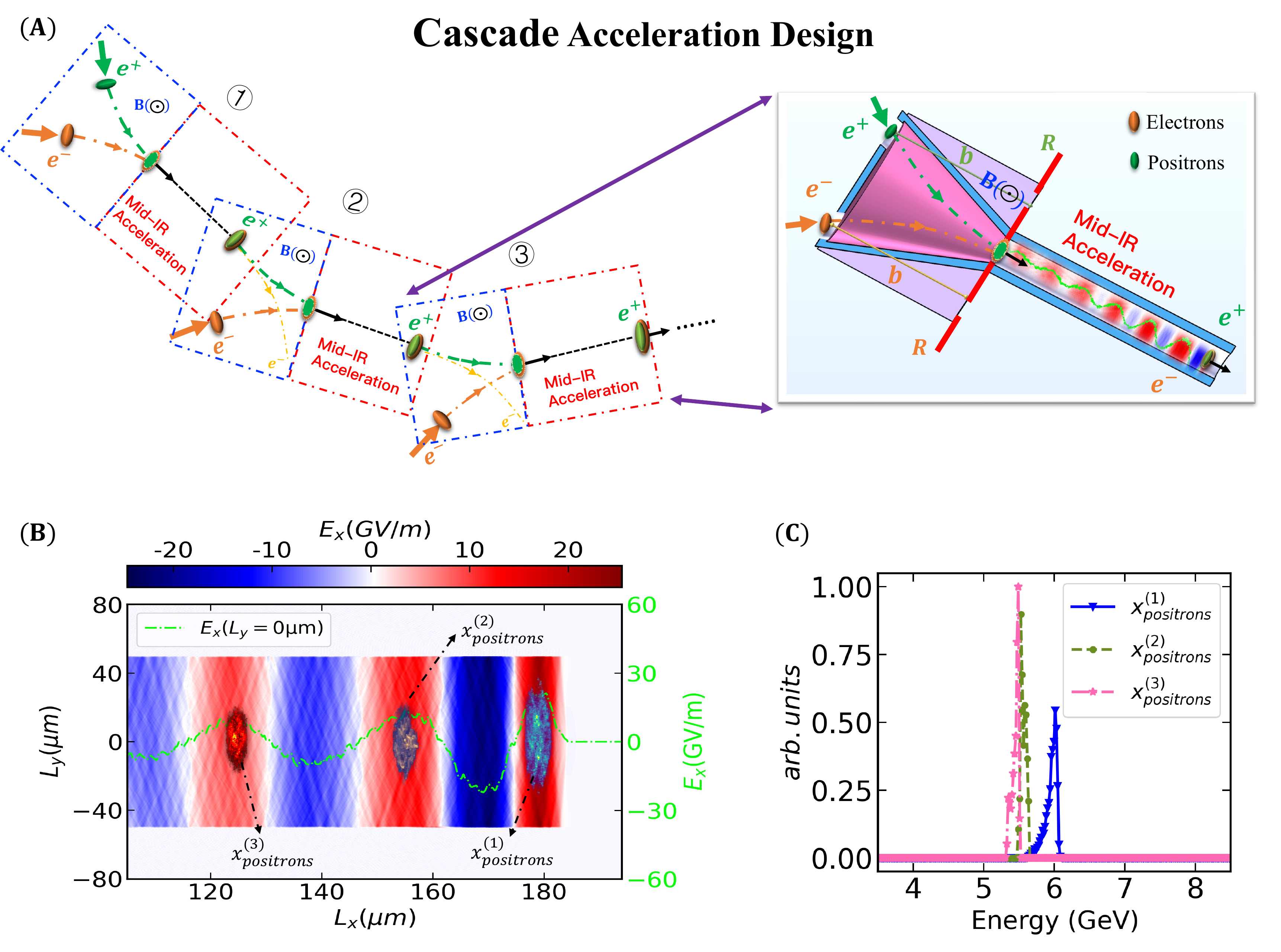}
    \caption{The schematic diagram of the cascade acceleration of a single positron bunch, and the series acceleration of multiple positron bunches: ({\bf A}) The cascade design for the injection of electron beams and a positron bunch. A cone-shaped coupling tube surrounded by a magnetic field. R is the radius of deflection; ({\bf B}) The mid-IR acceleration field accelerating three positron bunches, $x_{positrons}^{(1)}$, $x_{positrons}^{(2)}$, and $x_{positrons}^{(3)}$, simultaneously, with the initial positions of $178$ $\mathrm{\mu m}$, $156$ $\mathrm{\mu m}$, and $126$ $\mathrm{\mu m}$, respectively; and ({\bf C}) The final energy spectrum of the positron bunches with the initial energy of $5$ $\mathrm{GeV}$ at $t=140$ $\mathrm{ps}$. }
    \label{Figure5}
\end{figure}

\section{Conclusion}

In this paper, we proposed a novel scheme for the acceleration of positron bunches using high-intensity, continuous mid-IR field. The injected positron bunch could gain energy of $1$ $\mathrm{GeV}$ after accelerating for a distance of $\sim4.2$ $\mathrm{cm}$, and could maintain a low energy spread of $<2\%$. The energy transfer efficiencies ($\eta$) for two positron bunches with a charge of $0.111$ $\mathrm{nC}$ and $0.560$ $\mathrm{nC}$ were observed to be $\sim5\%$ and $\ 22\%$, respectively. In the uniform mid-IR acceleration field, the transverse field in the $y$-direction ($E_y-cB_z$) weakly focused on the positron bunch within the first $21$ $\mathrm{ps}$, post which it began weakly defocusing, till it finally reduced to zero due to the energy loss and beam break of the driving electron beam. The ease of performing the experiment was increased due to the excellent tolerance of the positron bunch injection deviation. Through several 2-dimensional PIC simulations, the transverse spatial misalignment of the positron injection is allowed to be a few microns if $10\%$ relative loss of the energy transfer efficiency is accepted, where the accelerated energy and the energy spread are hardly affected. The novel high-efficiency scheme had a high acceleration gradient and a stable acceleration structure, allowing for a stable, efficient, and continuous acceleration of multiple positron bunches simultaneously.

\section{Methods}
The proposed scheme is confirmed by cylindrical geometry with a decomposition in azimuthal modes using the SMILEI code\cite{D2018}, which is an open-source, user-friendly, high-performance and multi-purpose electromagnetic particle-in-cell (PIC) code for plasma simulation. The grid coordinates are two-dimensional $(x,r)$, while particle coordinates are expressed in the three-dimensional Cartesian frame $(x,y,z)$. A moving window of 51.2 $\mathrm{\mu m}$ in the $x$ direction and 100 $\mathrm{\mu m}$ in the $r$ direction was employed, which was sampled by 512 cells in the $x$ direction and 400 cells in the $r$ direction. Electromagnetic boundary conditions were implemented using the Perfectly Matched Layer (PML) to absorb fields at the boundary. Additionally, a binomial filter module was applied to the currents to reduce noise.
The plasma critical number density $n_c$ is $2.82\times 10^{25}$ $\mathrm{m^{-3}}$ for the $\lambda=2\pi$ $\mathrm{\mu m}$. The density of the dense plasma is $10^{28}$ $\mathrm{m^{-3}}$ with the radius of 50 $\mathrm{\mu m}$ and the thickness of 30 $\mathrm{\mu m}$. The initial energy of the driving electron bunch is 1 $\mathrm{GeV}$, with a 3.402 $\mathrm{nC}$ beam charge. The initial momentum is Maxwell-Juttner distribution with temperature 0.1 $\mathrm{m_ec^2}$ in the x directron and 0.001 $\mathrm{m_ec^2}$ in the y and z direction. The density profile of the electron bunch is a Gaussian distribution in the x and r direction, $n_e=n_{e0}e^{-(x-x_0)^2/\sigma_{xe}^2}\cdot e^{-r^2/\sigma_{re}^2}$, where $n_{e0}=4\times10^{24}$ $\mathrm{m^{-3}}$, $x_0=5$ $\mathrm{\mu m}$, longitudinal rms-length $\sigma_{xe}=3.3$ $\mathrm{\mu m}$, transverse rms-length $\sigma_{re}=26$ $\mathrm{\mu m}$. The witness positron bunch has an initial energy of 5 $\mathrm{GeV}$ or 1 $\mathrm{GeV}$ and a beam charge of 0.111 $\mathrm{nC}$. The density of the positron bunch is $n_p=3\times10^{23}$ $\mathrm{m^{-3}}$ with longitudinal rms-length  $\sigma_{xp}=2$ $\mathrm{\mu m}$, transverse rms-length  $\sigma_{rp}=22$ $\mathrm{\mu m}$, and the initial positron is 33.5 $\mathrm{\mu m}$. The initial energy spread, angular distribution and the normalized emittances of the positron bunch is neglected. There are 20 macro-particles per cell (PPC) for the driving electron bunch and the witness positron bunch, and 4 PPC for the plasma species. The results have been compared with PPC=50 for particles bunch and PPC=4/100/200/400 for the plasma particles of micro-tube, which are almost no difference. The analysis of transverse off-axis injection of positron bunch and series acceleration of multiple positron bunches are simulated with 2-dimensions PIC. A moving window of 204.8 $\mathrm{\mu m}$ ($L_x$) $\times$ 160 $\mathrm{\mu m}$ ($L_y$) is used and sampled by 2048 $(x)$ $\times$ 640 $(y)$ cells. 

\section*{Acknowledgments}
{\bf Funding:} This work is supported in part by National Natural Science Foundation
of China (11655003); Innovation Project of IHEP (542017IHEPZZBS11820,
542018IHEPZZBS12427); the CAS Center for Excellence in Particle Physics
(CCEPP); IHEP Innovation Grant (Y4545170Y2); Chinese Academy of Science
Focused Science Grant (QYZDY-SSW-SLH002); Chinese Academy of Science
Special Grant for Large Scientific Projects (113111KYSB20170005);
National 1000 Talents Program of China; the National Key Research
and Development Program of China (No.2018YFA0404300);
National Natural Science Foundation of China (Grant No. 11975252);
Youth Innovation Promotion Association CAS (No. 2021012).

\noindent{\bf Author contributions:} M.Y.S. contributed to investigation, validation, formal analysis, data curation, simulation, writing—original draft. Y.S.H. contributed to conceptualization, methodology, formal analysis, supervision, project administration, funding acquisition, writing—review and editing. M.Q.R. contributed to formal analysis. B.F.S. contributed to formal analysis.  Z.L.X. helped in formal analysis. T.P.Y. helped in formal analysis, writing—review and editing. X.F.W. contributed to formal analysis. Y.C. contributed to formal analysis.

\noindent{\bf Competing interests:}The authors declare no conflicts of interest.

\noindent{\bf Data and materials availability:} This manuscript has no associated data or the data will not be deposited.

\section*{References}
\begin{quote}

\end{quote}


\begin{enumerate}
\bibitem{L1979}L. Marchut, C. J. Mcmahon, Electron and positron spectroscopies in materials science and engineering. {\it Electron Positron Spectroscopies in Materials Science Engineering\/}, 1–33 (1979).
\bibitem{G2014}Guessoum, Nidhal, Positron astrophysics and areas of relation to low-energy positron physics. {\it European Physical Journal D\/} 68, 1-6 (2014).
\bibitem{H2006}K. R. Hogstrom, P. R. Almond, Review of electron beam therapy physics. {\it Physics in Medicine, Biology\/} 51, R455 (2006).
\bibitem{Gschwendtner}E. Gschwendtner, P. Muggli, Plasma wakefield accelerators. {\it Nature Reviews Physics\/} 1, 246-248 (2019).
\bibitem{Jiao}X. J. Jiao, J. M. Shaw, T. Wang et al, A tabletop, ultrashort pulse photoneutron source driven by electrons from laser wakefield acceleration. {\it Matter and Radiation at Extremes\/} 2, 296-302 (2017).
\bibitem{Tajima}T. Tajima, J. M. Dawson, Laser electron accelerator. {\it Physical Review Letters\/} 43, 267 (1979).
\bibitem{Katsouleas}Katsouleas, Thomas, Electrons hang ten on laser wake. {\it Nature\/} 431, 515-516 (2004).
\bibitem{Faure}J. Faure, Y. Glinec, A. Pukhov et al, A laser–plasma accelerator producing monoenergetic electron beams. {\it Nature\/} 431, 541-544 (2004).
\bibitem{Mangles}S. P. Mangles, C. D. Murphy, Z. Najmudin et al, Monoenergetic beams of relativistic electrons from intense laser–plasma interactions. {\it Nature\/} 431, 535-538 (2004).
\bibitem{Geddes}C. G. R. Geddes, C. Toth, J. Van Tilborg et al, High-quality electron beams from a laser wakefield accelerator using plasma-channel guiding. {\it Nature\/} 431, 538-541 (2004).
\bibitem{WangJ} Wang, J., Zeng, M., Li, D., Wang, X., Lu, W., Gao, J, Injection induced by coaxial laser interference in laser wakefield accelerators. {\it Matter and Radiation at Extremes\/} 7, 054001 (2022).
\bibitem{DaiYN}Dai, Y. N., Shen, B. F., Li, J. X., Shaisultanov, R., Hatsagortsyan, K. Z., Keitel, C. H., Chen, Y. Y. , Photon polarization effects in polarized electron–positron pair production in a strong laser field. {\it Matter and Radiation at Extremes\/} 7, 014401 (2022).
\bibitem{2016}T. J. Xu, B. F. Shen et al, Ultrashort megaelectronvolt positron beam generation based on laser-accelerated electrons. {\it Physics of Plasmas\/} 23, 033109 (2016).
\bibitem{Zhu} X. L. Zhu, M. Chen, T. P. Yu et al, Collimated GeV attosecond electron–positron bunches from a plasma  driven by 10 PW lasers. {\it Matter and Radiation at Extremes\/} 4, 014401 (2019).
\bibitem{L2007}K. V. Lotov, Acceleration of positrons by electron beam-driven wakefields in a plasma. {\it Physics of Plasmas \/}14, 023101 (2007).
\bibitem{D2000}C. Du, Z. Xu, Positron acceleration by a laser pulse in a plasma. {\it Physics of Plasmas\/} 7, 1582-1585 (2000).
\bibitem{H2003}H. Hasegawa, S. Usami, Y. Ohsawa, Positron acceleration to ultrarelativistic energies by a shock wave in a magnetized electron–positron–ion plasma. {\it Physics of Plasmas\/} 10, 3455-3458 (2003).
\bibitem{W2008}X. Wang, R. Ischebeck et al, Positron Injection and Acceleration on the Wake Driven by an Electron Beam in a Foil-and-Gas Plasma. {\it Physical Review Letters\/} 101, 124801 (2008).
\bibitem{V2014}J. Vieira, J.T. Mendonça, Nonlinear laser driven donut wakefields for positron and electron acceleration. {\it Physical Review Letters\/} 112, 215001(2014).
\bibitem{Schroeder1999}C. B. Schroeder, D. H. Whittum, J. S. Wurtele, Multimode analysis of the hollow plasma channel wakefield accelerator. {\it Physical Review Letters\/} 82, 1177 (1999).
\bibitem{Schroeder2013}C. B. Schroeder et al, Beam loading in a laser-plasma accelerator using a near-hollow plasma channel. {\it Physics of Plasmas\/} 20, 123115 (2013).
\bibitem{Gessner2016}S. Gessner, E. Adli et al, Demonstration of a positron beam-driven hollow channel plasma wakefield accelerator. {\it Nature communications\/} 7, 1-6 (2016).
\bibitem{Luwei}S. Zhou, J. Hua et al, High efficiency uniform wakefield acceleration of a positron beam using stable asymmetric mode in a hollow channel plasma. {\it Physical Review Letters\/} 127, 174801(2021).
\bibitem{Y2014}L. Yi, B. Shen et al, Positron acceleration in a hollow plasma  up to TeV regime.  {\it Scientific Reports\/} 4, 4171 (2014).
\bibitem{Xu2020}Z. Xu, L. Yi et al, Driving positron beam acceleration with coherent transition radiation. {\it Communications Physics\/} 3, 191 (2020).
\bibitem{Xu2023}Xu, Z., Shen, B., Si, M., Huang, Y. S.. Positron acceleration by terahertz wave and electron beam in plasma channel. {\it New Journal of Physics\/} (2023).
\bibitem{ZhuXL} Zhu, X. L., Liu, W. Y., Weng, S. M., Chen, M., Sheng, Z. M., Zhang, J, Generation of single-cycle relativistic infrared pulses at wavelengths above 20 $\mathrm{\mu m}$ from density-tailored plasmas. {\it Matter and Radiation at Extremes} 7, 014403 (2022).
\bibitem{Si}Meiyu Si, Yongsheng Huang, Manqi Ruan, Baifei Shen, Zhangli Xu, Tongpu Yu, Xiongfei Wang, and Yuan Chen, "Relativistic-guided stable mode of few-cycle 20 $\mathrm{\mu m}$ level infrared radiation," {\it Optics. Express\/} 31, 40202-40209 (2023).
\bibitem{Londrillo}P. Londrillo, C. Gatti, M. Ferrario, Numerical investigation of beam-driven PWFA in quasi-nonlinear regime. {\it Nuclear Instruments and Methods in Physics Research Section A: Accelerators, Spectrometers, Detectors and Associated Equipment\/} 740, 236-241 (2014).
\bibitem{Massimo}F. Massimo, A. Marocchino, A. R. Rossi, Electromagnetic self-consistent field initialization and fluid advance techniques for hybrid-kinetic pwfa code architect. {\it Nuclear Instruments and Methods in Physics Research Section A: Accelerators, Spectrometers, Detectors and Associated Equipment\/} 829, 378-382 (2016).
\bibitem{Marocchino}A. Marocchino, E. Chiadroni et al, Design of high brightness Plasma Wakefield Acceleration experiment at SPARC-LAB test facility with particle-in-cell simulations. {\it Nuclear Instruments and Methods in Physics Research Section A: Accelerators, Spectrometers, Detectors and Associated Equipment\/} 909, 408-413 (2018).
\bibitem{D2018}J. Derouillat et al, SMILEI: A collaborative, open-source, multi-purpose particle-in-cell code for plasma simulation. {\it Computer Physics Communications\/} 222, 351-373 (2017).
\end{enumerate}
\end{document}